\documentclass[twocolumn, pra, nofootinbib]{revtex4-1}
\usepackage{amsmath, amsfonts, amssymb, array, mathtools}
\usepackage{mathrsfs}
\usepackage{xcolor}
\usepackage{graphicx}

\usepackage{slashed}
\usepackage{hyperref}

\hypersetup{
  colorlinks   = true, 
  urlcolor     = blue, 
  linkcolor    = blue, 
  citecolor   = blue 
}

\begin{document}
\title{Nonclassical states of light in a nonlinear Michelson interferometer}

\author{Bijoy John Mathew} 
\affiliation{Indian Institute of Science Education and Research Thiruvananthapuram, Maruthamala PO, Thiruvananthapuram, Kerala, India 695551}
\email{bijoy13@iisertvm.ac.in}

\author{Anil Shaji}
\affiliation{Indian Institute of Science Education and Research Thiruvananthapuram, Maruthamala PO, Thiruvananthapuram, Kerala, India 695551}

\begin{abstract}
Nonlinear quantum metrology schemes can lead to  faster than Heisenberg limited scalings for the measurement uncertainty. We study a Michelson interferometer embedded in a Kerr medium [Luis and Rivas, Phys.~Rev.~A {\bfseries 92}, 022104 (2015)] that leads to non-linear, intensity dependent phase shifts corresponding to relative changes in the lengths of its two arms. The quantum Cramer-Rao bound on the minimum achievable measurement uncertainties is worked out and the requirements, in practice, to saturate the bound are investigated. The choice of input state of light into the interferometer and the read out strategy at the output end are discussed. The ideal, non-classical states of light that must be used to saturate the bound are found to be highly susceptible to photon loss noise. We identify  optimal states at each noise level that are both resilient to noise and capable of giving the enhanced sensitivities and discuss practical implementations of the interferometry scheme using such states. 
\end{abstract}

\allowdisplaybreaks
\maketitle
\section{Introduction}\label{intro}

The ability to make precise measurements is a crucial requirement in experimental physics as we continue to explore the boundaries of our knowledge of the physical universe. Detection of gravitational waves via measurement of optical phase generated by a passing wave in an interferometer is one such area where the quantum theory of metrology plays a vital role~\cite{Adhikari2014,Ligo:2011dj,Grote:2013ej,Aasi:2013jb}. To measure a physical parameter, a suitable probe whose state evolves in a manner that depends on the value of the parameter is used. The value of the parameter is estimated by a read-out of the final state of the probe. If the probe is made of $N$ independent \textcolor{black}{identical} classical systems, each of which is capable of giving an estimate of the parameter, then it is well known that the measurement uncertainty in the parameter is shot-noise limited and scales as  $1/\sqrt{N}$~\cite{Giovannetti2011}. If the measuring device is an interferometer in which the value of the parameter to be measured is imprinted as a phase shift, then $N$ stands for the number of photons in the laser light traversing the interferometer. The  $1/\sqrt{N}$ scaling of the measurement uncertainty is also known as the standard quantum limit (SQL).

Quantum metrology refers to the scenario in which the quantum features of the $N$ physical systems that make up the probe can be taken advantage of for achieving better precision in the measurement. Since all physical systems are quantum mechanical in nature to the best of our knowledge, quantum metrology also addresses the question of the fundamental limits on the measurement precision given the constraints on the resources available for performing a measurement. The number $N$ of systems that make up the probe is effectively a placeholder for the specific physical resource like energy, momentum, time, particle number, etc., that is relevant for the measurement being considered. It is known that by utilising the quantum features of the probe units, one can go beyond the shot noise limited scaling and get to the so called Heisenberg limited scaling of $1/N$ for the measurement uncertainty. In the case of interferometry,  several schemes have been proposed that theoretically approach the Heisenberg limit by using nonclassical states of light such as maximally entangled N00N states, squeezed states, etc~\cite{Caves1981,Yurke1986,Bollinger1996,Lee2002,Giovannetti2006,Higgins2007,Berry2009,Pezze2018}. Some of these quantum metrology protocols have been implemented in gravitational wave interferometers~\cite{Ligo:2011dj,Grote:2013ej,Aasi:2013jb}.

One of the assumptions that goes into obtaining the $1/N$ scaling is that each of the probe units couple independently to the parameter being measured. In other words, the coupling Hamiltonian that generates the parameter dependent evolution has the form 
\begin{equation}
	\label{hamil1}
	H_{\rm probe} = \phi \sum_{j} h_{j}, \quad j=1,\ldots, N,
\end{equation}
where $\phi$ is the measured parameter and $h_{j}$ are operators on each of the units of the quantum probe.  However, if an effective nonlinear coupling can be engineered between the probe units and the parameter, then a faster scaling than $1/N$ for the measurement uncertainty can be achieved. \textcolor{black}{Nonlinear transformations leading to a metrological advantage was considered in the context of interferometry in~\cite{Luis2004} and this idea was developed further in~\cite{Beltran2005}. Using nonlinear generators of parameter dependent evolution in the context of metrology in general to obtain faster than $1/N$ scaling was first discussed in~\cite{Boixo2007} and later in~\cite{Luis2007}. Possible means of implementing such measurement schemes using atomic systems, Bose-Einstein condensates etc.~were considered in~\cite{Rey2007,Choi2008,Boixo2009,Napolitano2010,Tacla2010} while special cases like restricting the input product states and separable states were considered in~\cite{Roy2008,Boixo:2008bca,2Boixo2009,Rivas2010}. Laboratory implementations of such schemes have also been attempted~\cite{,Napolitano2011,Sewell2014,Toth2014}}. Whether the enhanced scaling can be thought of as an improvement over Heisenberg limited scaling or if it is the resource counting that has to change with the introduction of the nonlinear coupling is a matter of debate. For simplicity and clarity, we steer clear of this question and state directly that we will be referring to the enhanced scaling (anything faster than $1/N$) as super-Heisenberg scaling.  

\textcolor{black}{The best possible scaling for the measurement precision for a given probe Hamiltonian can be obtained in terms of the operator semi-norm, $\|H_{\rm probe}\|$, of the coupling Hamiltonian (linear or nonlinear) that generates the parameter dependent evolution of the probe~\cite{Boixo2007}. For any initial state of the probe and coupling Hamiltonian, $H_{\rm probe}$, the best possible precision is given by the quantum Cramer-Rao bound~\cite{Helstrom1967,Helstrom1969,Braun1994,Braun1996}. Achieving the bound that saturates the best possible scaling possible determined by   $\|H_{\rm probe}\|$ usually involves using highly entangled initial states of the probe. These states typically are not at all robust against external noise and decoherence. However, with nonlinear couplings in $H_{\rm probe}$, it has been shown that faster than $1/N$ scaling can still be obtained using semiclassical or product states of the probe even if the best possible scaling is not achieved. Such states tend to be more robust against decoherence~\cite{Beltran2005,Rivas2010,Boixo:2008bca}}. In the context of interferometry, effective nonlinearities that can potentially lead to super-Heisenberg metrology have been demonstrated in light propagation through Kerr-nonlinear media~\cite{Pace1993,Rehbein2005,Khalaidovski2009}.  Recently, Luis and Rivas~\cite{Luis2015} introduced a quantum detection scheme using a Michelson interferometer embedded in gas with Kerr nonlinearity as a candidate system that can demonstrate super-Heisenberg scalings. Coherent light pulses are used as input states in the scheme and light pulse duration was treated as a new variable that comes into play in the measurement.  

In this Paper, we use the same interferometry scheme as Luis and Rivas but generalise its scope substantially by considering nonclassical states of light as inputs instead of the coherent states considered in~\cite{Luis2015}.  We recap the nonlinear interferometer in Sec.~\ref{scheme} and in Sec.~\ref{optimal}  we compute the quantum Cramer-Rao bound as applicable to this case and find the best, theoretically possible, scaling for the measurement uncertainty for such an interferometer configuration with and without photon loss noise. We examine the actual performance of the interferometry scheme against the quantum Cramer-Rao bound in Sec.~\ref{detection}. To achieve the theoretical bounds on the measurement uncertainty in practical implementations, both the input state of light into the interferometer as well as the final read-out of the output state can both be optimised. We explore both of these optimisations for the cases with and without photon loss. Finally, in Sec.~\ref{optimalstate} we identify an optical nonclassical state light that can be used in such an interferometer which leads to the best possible measurement precision while at the same time being resilient to photon loss noise. Our conclusions are in Sec.~\ref{conclude}.

\section{The nonlinear interferometer}\label{scheme}

\begin{figure}
	\resizebox{8.5cm}{7cm}{\includegraphics{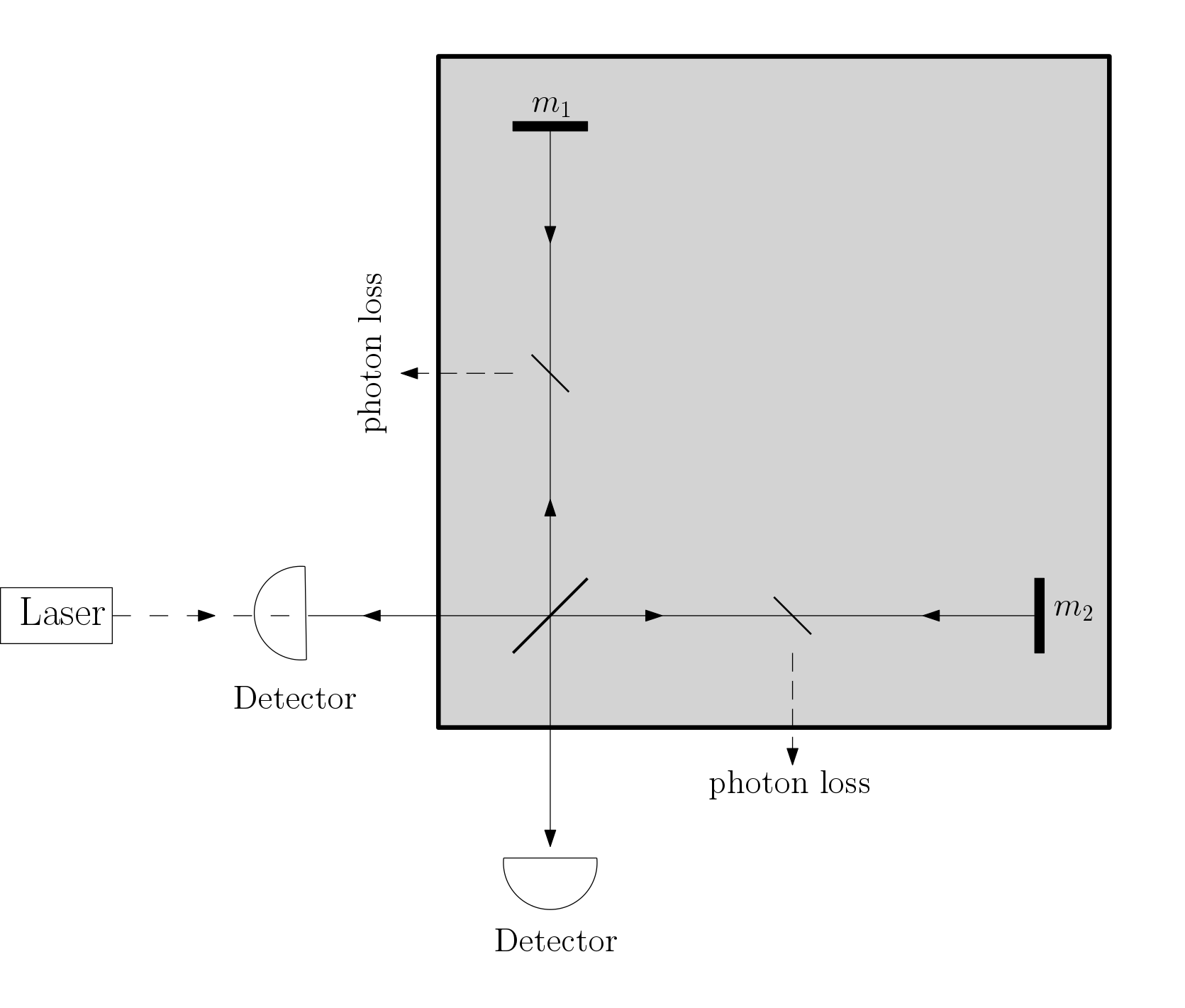}}
\caption{Schematic diagram for a Michelson interferometer embedded in Kerr media~\cite{Luis2015}. Photon loss can be modelled as beam splitters in each arm that remove a fraction of the photons in that arm \label{fig:scheme}}
	
\end{figure}

Luis and Rivas considered a Michelson interferometer embedded in a gas exhibiting a Kerr nonlinearity. Classical light pulses (coherent states) are injected into one of the input ports of the interferometer while the other input is kept empty with only the quantum vacuum entering that port. A passing gravitational wave signal, $x$, manifests itself as anti-correlated length changes of the two arms of the interferometer given by $l_1 = l_0 -x/2$, $l_2 = l_0 + x/2$, where $l_{0}$ is the original length of each arm. The changes in length are computed from the changes in the relative phase of the light in the two arms that, in turn, is detected by counting the number of photons coming out at each output port of the interferometer~\cite{Adhikari2014}. A schematic diagram of the interferometer is given in Fig.~\ref{fig:scheme}.
The refractive index of the Kerr medium that envelopes the interferometer has an intensity-dependent term leading to the nonlinearity in the relevant Hamiltonian operator that generates the relative phase between the two arms. The refractive index of the gas can be expressed as,
\begin{equation}
n=n_0 + \tilde{n}I = n_0(1+\chi N),
\end{equation}
where $n_0$ is the linear index, $N$ is the number of photons circulating in the interferometer, $\tilde{n}$ is a coefficient related to the third-order susceptibility $\chi^{(3)}$ of the Kerr medium that quantifies the nonlinearity of the refractive index and $\chi$, which is proportional to $\chi^{(3)}$ captures the nonlinear phase shift per photon.

Photons that enter the interferometer arms from the beam splitter pick up the phase shift while propagating through the arms, bounce back from the mirrors, rejoin at the beam splitter and are finally detected at the output ports. Light propagation is described in terms of the internal modes of the interferometer, ${a_1, a_2}$, by the unitary operator $U=U_1U_2$ where
\begin{equation}
U_j = e^{\iota\phi_j \hat{G}_j}, \qquad \hat{G}_j = \hat{N}_j + \frac{\chi}{2}\hat{N}^2_j. \label{eqn:unitary}
\end{equation}
in the equation above, $\phi_j = \bar{k}l_j$, $j = 1,2$ where $\bar{k}$ is the wave number of the light used and $\hat{N}_j = a^{\dagger}_ja_j$ is the photon number operator for each arm. The two detectors at the output ports of interferometer record the difference in the number of photons detected at each port, allowing us to infer the relative phase difference between the arms of the interferometer. In the following, instead of applying the beam-splitter transformation on the state of light that has traversed through the interferometer so as to find the state at the output ports of the interferometer, we take the measurement operator corresponding to the difference in photo-counts at the output ports, $M' =(b_{2}^{\dagger}b_{2}^{\vphantom{\dagger}} - b_{1}^{\dagger}b_{1}^{\vphantom{\dagger}})$, with ${b_1, b_2}$ being the external modes of the interferometer, and apply on it the (inverse) beam-splitter transformation to find its form immediately before the returning light hits the beam splitter. This choice makes most of our calculations significantly simpler. The corresponding measurement operator in terms of the internal modes is,
\begin{equation}
	\label{eq:meas1} 
	M = i(a_2^\dagger a_1 - a_1^\dagger a_2).
\end{equation}
When a coherent state is injected into one of the input ports of the interferometer as in~\cite{Luis2015}, the measurement uncertainty, as quantified by the standard deviation of the estimate of $x$ obtained from the measurement of $M$, is found to scale as $1/N^{3/2}$. 

The choice of the input state considered by Luis and Rivas is typical of interferometers, including those typically used for gravitational wave detection. The read-out of the state of light at the output end also conforms to such a typical configuration. However, these choices do not necessarily represent ones that correspond to the absolute minimum possible measurement uncertainty that can be achieved using such an interferometer configuration with a nonlinear phase shift. We explore the minimum possible uncertainty in the next couple of sections. 

\section{Theoretical bound on the measurement uncertainty}\label{optimal}

The super-Heisenberg scaling of $1/N^{3/2}$ is not the best possible scaling for the interferometry scheme considered here. The theoretical bound on the measurement uncertainty is given by the quantum Cramer-Rao bound (QCRB) that in turn is phrased in terms of the quantum Fisher information~\cite{Helstrom1967,Helstrom1969,Braun1996,Braun1994}. The quantum state of light in the interferometer is effectively the quantum probe in this case. The QCRB captures the idea that the measurement uncertainty is inversely proportional to the rate of change of the state of the probe in its Hilbert space in response to changes in the measured parameter (phase $\phi$ in the present case). \textcolor{black}{As mentioned earlier,} the bound on the measurement uncertainty given by the QCRB depends only on the input state of light and the parameter dependent dynamics that it goes through within the interferometer. The bound is independent of the particular read-out strategy that is employed to determine the relevant information from the final state of the probe that then leads to the estimate of the parameter of interest. Once the QCRB is known, one still has to find the best read-out strategy on the output probe state that either saturates the bound or comes close to it given the practical limitations of a laboratory interferometer setup. Before considering these questions, we first compute the QCRB with and without photon loss noise for the nonlinear interferometry scheme. 

\subsection{Ideal case: No photon loss}

We write the part of the probe Hamiltonian that imprints the relative phase $\phi$ on the light in the interferometer as $H_\phi = \phi H$. The change in the state of light, $\rho$, with respect to changes in $\phi$ is generated by the operator $H$ as
\begin{equation}
	\label{eq:dyn1}
	\rho' = \frac{\partial\rho}{\partial\phi} = -i[H,\rho]
\end{equation}
The quantum Cramer-Rao bound for the measurement uncertainty in estimating $\phi$ is then given in terms of the quantum Fisher information, $\mathcal{F}$, as~\cite{Braun1996}
\begin{equation}
	(\delta \phi)^{2} \geq \frac{1}{\nu\mathcal{F}} = \frac{1}{\nu \left\langle \mathcal{L}^2_{\rho}(\rho') \right\rangle },
\end{equation}
\textcolor{black}{where $\nu$ is the number of times the measurement is repeated and $\mathcal{L}$ is the symmetric logarithmic derivative operator. The measurement uncertainty decrease obtained via the repetition factor $\nu$ is not important to our framework, so we will ignore it for the rest of the analysis}. Now, $\mathcal{L}_\rho (\rho')$ can be expressed in the basis $\left\lbrace \left| j\right\rangle\right\rbrace $ that diagonalizes $\rho$, as~\cite{Braun1994}
\begin{equation}
	\label{SLD}
\mathcal{L}_{\rho}(\rho')= \sum_{p_j + p_k \neq0} \frac{2}{p_j + p_k}\rho_{jk}'\left| j\right\rangle \left\langle k\right|,
\end{equation}
where $\rho_{jk}' = \left\langle j\right| \rho' \left| k\right\rangle$ \textcolor{black}{and $\left\lbrace p_j\right\rbrace $ are the eigenvalues of $\rho$}. The quantum Fisher information then can be written using Eq.~(\ref{eq:dyn1}) as,
\begin{multline}
\mathcal{F} = 4\left[ \sum_{j} p_j \Big(  \left\langle j\right| H^2 \left| j\right\rangle - \big| \left\langle j\right| H \left| j\right\rangle\big|^2 \Big) \right. \\
\left. -\sum_{j \neq k} \frac{4p^2_jp_k}{(p_j + p_k)^2} \big|\left\langle j\right| H \left| k\right\rangle\big|^2 \right] \label{eqn:fisher}
\end{multline}
From equation \eqref{eqn:unitary}, assuming $\phi_2 = \phi/2$ and $\phi_1 = -\phi/2$, the generator of the relative phase  $\phi$ for the interferometry scheme can be written as,
\begin{equation}
H = \frac{\hat{G_2}-\hat{G_1}}{2}.
\end{equation}

A quantum state of light that is constructed by superposing eigenstates of $H$ that correspond to its largest and smallest eigenvalues is a good candidate for the optimal probe state that changes rapidly with small changes in $\phi$. Adding to this the requirement that the initial state of the probe that we consider also be one with a fixed photon number \textcolor{black}{would lead us to the 'N00N' state, 
\[ |\psi\rangle_{\rm N00N} =  \frac{\left| N, 0\right\rangle + \left| 0, N \right\rangle}{\sqrt{2}}. \]
However for reasons that would be clear in Sec.~\ref{detection}, we consider a more general state of the form,}
\begin{equation}
	\left| \psi\right\rangle  = \frac{\left| N-k, k\right\rangle + \left| k,N-k\right\rangle}{\sqrt{2}}, \quad k \ \in \ [0,N] \label{eqn:input}
\end{equation}
Note that $\left| \psi\right\rangle$ is the state entering the interferometer arms after the beam splitter and it is not the state that is injected into the input port(s). In order to obtain the state that has to be injected, the inverse beam splitter transformation may be applied. \textcolor{black}{Note that for N00N states, $k=0$}.  The state of light reaching the beam splitter after propagation through the interferometer arms is:
\begin{multline}
U\left| \psi\right\rangle = \frac{1}{\sqrt{2}} \left[ e^{i\phi_1\widetilde{G}_{N-k}} e^{i\phi_2\widetilde{G}_{k}} \left| N-k, k\right\rangle \right. \\
\left. +\  e^{i\phi_1\widetilde{G}_{k}} e^{i\phi_2\widetilde{G}_{N-k}} \left| k, N-k\right\rangle \right] \label{eqn:final}
\end{multline}
where $\widetilde{G}_n = n + \frac{\chi}{2}n^2$. Here we have assumed that the changes produced in the optical phase by sources other than the gravitational wave signal $x$ are negligible.

Since the state of light in the interferometer in the ideal case without noise is always a pure state, Eq.~\eqref{eqn:fisher} reduces to
\begin{align}
\label{eq:Fisher}
\mathcal{F} = 4(\Delta H)^2 &= (\widetilde{G}_{N-k} - \widetilde{G}_k)^2 \nonumber \\
&= \left( N-2k + \frac{\chi N^2}{2} - \chi Nk \right)^2
\end{align}
The Fisher information is maximum when $k=0$ so that all the terms with negative signs in the expression for ${\cal F}$ above are zero. This choice corresponds to the initial state being the N00N state. In this case ${\sqrt{\mathcal{F}}} \sim N^{2}(\chi/2 + 1/N)$. Since the relative phase $\phi$ is directly proportional to $x$, \textcolor{black}{($\phi = \bar{k} x$)}, the QCRB on the measurement uncertainty in estimating the change in length of the arms of the interferometer  is 
\begin{equation}
\label{crb1}
\delta x \sim \delta \phi \geq \frac{1}{\sqrt{\mathcal{F}}} \sim \frac{1}{\chi N^2}
\end{equation}
\textcolor{black}{As shown in ~\cite{Boixo2007}, $1/N^{2}$ scaling is the best that can be expected with the $\chi N^2/2$ (quadratic) nonlinearity considered here}. The QCRB tells us that using the optimal standard deviation $\Delta x$ for any configuration of the interferometer can scale at best as $1/N^{2}$, which beats the usual Heisenberg limit by a factor of $1/N$.

\subsection{With photon loss}

All practical interferometers are invariably affected by decoherence and the dominant source of noise is usually photon loss. This photon loss can be modeled by inserting beam splitters on both arms of the interferometer with transmissivities, $\eta_{a_1}$ and $\eta_{a_2}$, respectively, as shown in the schematic in Fig.~\ref{fig:scheme}. They can be inserted both before or after the phase is picked up by photons in the interferometer, as these operations commute with each other. The completely positive, trace preserving map that describes this model of photon loss is~\cite{DemDob2009}:
\begin{equation}
\textcolor{black}{\rho_n = \sum_{p,q = 0}^{\infty}K_{a_1,q}K_{a_2,p}\rho_0 K^{\dag}_{a_2,p}K^{\dag}_{a_1,q}} \label{eqn:loss}
\end{equation}
where
\begin{align}
K_{a_1,q} & = (1-\eta_{a_1})^{q/2}\eta_{a_1}^{a_1^{\dag}a_1/2}\frac{a_1^q}{\sqrt{q!}}, \nonumber \\
K_{a_2,p} & = (1-\eta_{a_2})^{p/2}\eta_{a_2}^{a_2^{\dag}a_2/2}\frac{a_2^p}{\sqrt{p!}}.
\end{align}
Setting $\eta_{a_1} = \eta_{a_2} = \eta$ and applying this map on the state in Eq.~\eqref{eqn:final}, we get
\begin{align}
\label{eq:noise1}
\rho_n &= \sum_{q,p=0}\left[ \frac{B_{pq}}{2}\sqrt{P^k_{ppqq}}\left| n^{N-k}_{q},n^{k}_{p}\right\rangle \! \left\langle n^{N-k}_{q},n^{k}_{p}\right| \right] \nonumber \\
&+ \sum_{q,p=0}\left[e^{-i\phi G(k)} \frac{B_{pq}}{2}\sqrt{P^k_{pqqp}} \left| n^{N-k}_{q}\!\!,n^{k}_{p}\right\rangle \! \left\langle n^{k}_{q},n^{N-k}_{p}\right| \right] \nonumber \\
&+ \sum_{q,p=0}\left[e^{i\phi G(k)} \frac{B_{pq}}{2}\sqrt{P^k_{qppq}} \left| n^{k}_{q},n^{N-k}_{p}\right\rangle \! \left\langle n^{N-k}_{q},n^{k}_{p}\right| \right] \nonumber \\
&+ \sum_{q,p=0}\left[ \frac{B_{pq}}{2}\sqrt{P^k_{qqpp}}\left| n^{k}_{q},n^{N-k}_{p}\right\rangle \! \left\langle n^{k}_{q},n^{N-k}_{p}\right| \right],
\end{align}
where  $B_{pq} = (1-\eta)^{p+q}\eta^{N-p-q}/p!q!$ with $G(k) = \widetilde{G}_{N-k}-\widetilde{G}_k$, and  $P^k_{abcd} = P(k,a)P(k,b)P(N-k,c)P(N-k,d)$,  with \textcolor{black}{$ P(i,j) = i!/(i-j)!$}. In Eq.~(\ref{eq:noise1}), the photon number in each of the Fock states is represented as $n^r_s = r-s$. The upper limits of the summations in the equation are the photon numbers of the states appearing in each term in the absence of photon loss ($k$ and $N-k$).

\begin{figure}[h]
	{\includegraphics[width = 8.5cm]{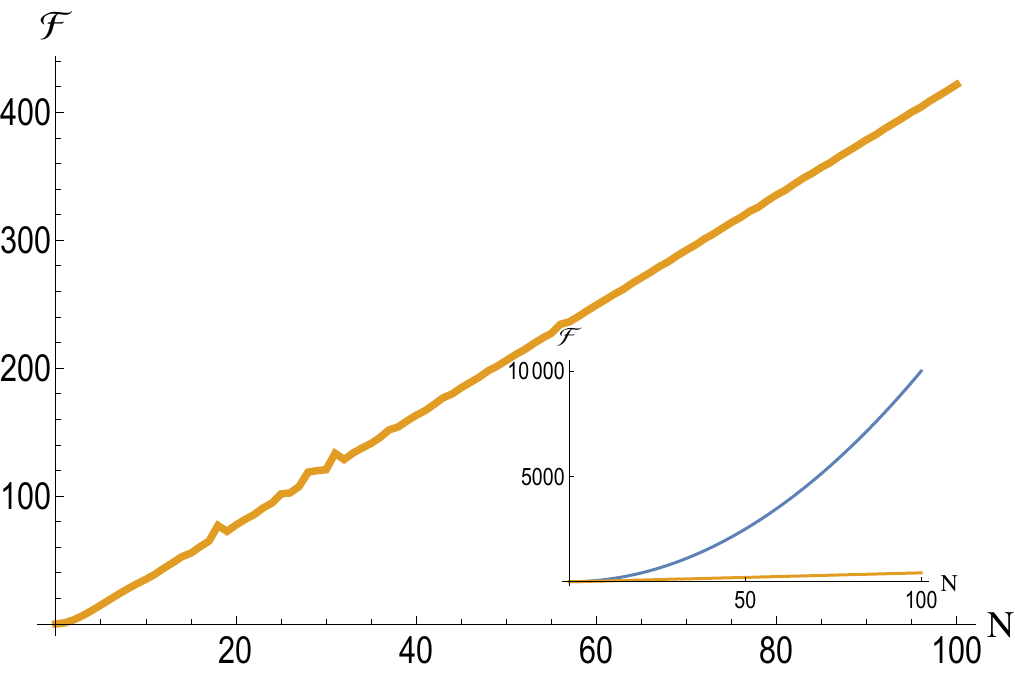}}
	\caption{Maximum quantum Fisher information as a function of $N$ for the state in Eq.~(\ref{eq:noise1}) with the transmission coefficient of the beam splitter modelling the photon loss set at $\eta = 0.9$. We see that the Fisher information scales linearly in $N$ and the super-Heisenberg scaling is lost when there is photon-loss. In the inset, a comparison of the quantum Fisher information with and without photon loss, setting $\eta = 1$ and $\eta = 0.9$ respectively is shown. \textcolor{black}{Note that $\mathcal{F}$, as given in Eq.~\eqref{eqn:fisher}, is plotted on the $y$-axis both in the main graph as well as in the inset. The (blue) curve in the inset corresponding to $\eta=1$ appears quadratic even if the behaviour is actually quartic in $N$ since the coefficient of the $N^4$ term in Eq.~\eqref{eq:Fisher} is proportional to $\chi \ll N$. The quartic behavior will be evident only for much larger values of $N$ than what we have considered here.}}
	\label{fig:lossqfi}
\end{figure}

The quantum Fisher information corresponding to the state with photon losses included was obtained numerically. We set the value of $\chi$ as $10^{-8}$ considering the giant nonlinearities reported in Ref~\cite{Molella2008}. The Quantum Fisher Information is computed for up to $N =100$ and for all cases corresponding to $k \in [0,N]$ and the maximum overall $k$ is taken. \textcolor{black}{For low values of the photon loss, the maximum occurs when $k$ takes values at the ends of the interval [0,N] corresponding to the N00N state. For larger losses, the maximum of the Quantum Fisher Information occurs for values of $k$ that are slightly away from the ends of the interval.} The plot of maximum of the Quantum Fisher Information versus $N$ is shown in Fig.~\ref{fig:lossqfi} for beam splitter transmittivities $\eta = 0.9$ and $\eta = 1$ (zero photon loss), respectively. It can be seen that with merely $10\%$ photon loss, the quadratic relationship between the maximum of the quantum Fisher information and the number of photons that goes beyond the Heisenberg limit in the absence of noise is quickly reduced to a linear one.

\section{Read-out of the probe}\label{detection}

The QCRB predicts that the best-case scenario for the nonlinear interferometer is an input state of type in Eq.~(\ref{eqn:input}). As mentioned earlier, the bound is independent of the read-out procedure used to gather the information about $\phi$ and $x$ from the final state of the probe. In the interferometer we consider, the read-out is photon counting at the output ports that corresponds to measurement of the operator $M$ given in Eq.~(\ref{eq:meas1}).  The signal $x$, through the phase difference $\Delta\phi = \phi_2 - \phi_1$ it creates between the interferometer arms as it passes by, would produce a shift in the mean value of $M$. We had concluded by looking at Eq.~(\ref{eq:Fisher}) that $k=0$ maximises the quantum Fisher information. However, it is easy to see from Eqs.~(\ref{eq:meas1}) and (\ref{eqn:final}) that the state of light at the output ports, in the case without noise, leads to a non-zero value for $\langle M \rangle$ only if  $k = (N+1)/2 $ or $k = (N-1)/2$. This highlights a problem that is often glossed over in quantum metrology. The nature of the quantum probe can lead to practical limitations on the types of read-outs that can be done on it. This means that in some cases, physical and practical considerations may limit one's ability to saturate the QCRB and it becomes meaningful to consider only the lowest achievable measurement uncertainty in the presence of the limitations on read-out~\cite{Jose:2013ia}.

In our case, for practical implementation, we therefore restrict to $k = (N+1)/2 $ or $k = (N-1)/2$. and we consider the input state (assuming $N$ to be odd),
\begin{equation}
\left| \psi\right\rangle  = \frac{1}{\sqrt{2}}\left[ \left| \frac{N-1}{2}, \frac{N+1}{2}\right\rangle + \left| \frac{N+1}{2}, \frac{N-1}{2}\right\rangle\right] \label{eqn:specific}
\end{equation}
After going through the interferometer arms, the state becomes
\begin{eqnarray}
\left| \psi\right\rangle  &= & \frac{1}{\sqrt{2}}\bigg[  e^{i \frac{\phi}{2} \big(G_{\frac{N+1}{2}}  -G_{\frac{N-1}{2}} \big)}\left| \frac{N-1}{2}, \frac{N+1}{2}\right\rangle  \nonumber \\
&& \qquad +\;  e^{-i \frac{\phi}{2} \big(G_{\frac{N+1}{2}}  -G_{\frac{N-1}{2}} \big)} \left| \frac{N+1}{2}, \frac{N-1}{2}\right\rangle\bigg] \qquad \label{eqn:specific2}
\end{eqnarray}
For this state we get, 
\begin{align}
\left\langle M \right\rangle &= \left( \frac{N+1}{2} \right) \sin\left[ \phi \left( \widetilde{G}_{\frac{N+1}{2}} - \widetilde{G}_{\frac{N-1}{2}}\right)\right] ,\nonumber \\
&= \left( \frac{N+1}{2} \right) \sin\left[ \bar{k}x \left( 1+\frac{\chi}{2}N\right)\right] , 
\end{align}
where $\bar{k}$ is the wave number of the (monochromatic) light used. For the state in Eq.~\eqref{eqn:specific2}, the variance, $\langle \Delta M^{2} \rangle = \langle M^2 \rangle  - \langle M \rangle^{2}$ is,
\begin{equation}
\langle \Delta M^{2}  \rangle  =\bigg( \frac{N+1}{2} \bigg)^{2} \bigg\{ \cos^{2} \left[ \bar{k}x \left( 1+\frac{\chi}{2}N\right)\right]+1\bigg\} - 2.
\end{equation}
The standard deviation in the estimate of $x$ is obtained using straightforward error propagation as
\begin{equation}
\bar{k}\Delta x = \Delta \phi = \frac{\sqrt{ \langle \Delta M^2 \rangle}}{| d\left\langle M \right\rangle/dx |}
\end{equation}
leading to
\begin{equation}
\textcolor{black}{\Delta x \simeq  \frac{2 \sqrt{2}}{\chi N},}
\end{equation}
for large $N$ and $x \rightarrow 0$  indicating that despite the nonlinearity in the interferometer, the restriction on the read-out procedure constrains the scaling of the measurement uncertainty to the Heisenberg limited one. However, this does not mean that the nonlinearity is not useful. If the nonlinear medium were absent, the minimum measurement uncertainty scales in this case as $1/\sqrt{N}$ when using the $N$ photons as independent probes of $x$. An improvement by a factor $1/\chi \sqrt{N}$ is therefore provided by the nonlinearity in this case.

\subsection{Multi-photon coincidence as read-out}

It is known that multi-photon coincidence detection has to be implemented at the output end if the full advantage provided by N00N states in the noiseless case is to be obtained~\cite{Jones:2009dm,Afek:2010fu,Israel:2014ii}. These read-outs, however, are typically extremely challenging to implement. \textcolor{black}{A few recent attempts towards implementing multi-photon coincidence detection can be found in~\cite{Branning2011,Zhu2018}}. We consider $m$-photon coincidence measurements at the output end described by the operator,
\begin{equation}
	\label{mphoton}
	 M_{m} = i \big[ (a_{1}^{\dagger})^{m} (a_{2}^{\vphantom{\dagger}})^{m} - (a_{1}^{\vphantom{\dagger}})^{m} (a_{2}^{\dagger})^{m}  \big],
\end{equation}
where we have written $M_{m} $ as it appears before the final beam splitter transformation for simplicity. 

When the state of light just before the last beam splitter of the interferometer is given by Eq.~(\ref{eqn:final}), we obtain, 
\[ \langle  M_{m} \rangle = C_{N,m} \sin \Big[  \Big(m+\frac{\chi}{2} Nm \Big) \phi \Big], \]
where we have chosen $k=(N-m)/2$ and 
\[ C_{N,m}  =  \frac{\big(\frac{N+m}{2} \big) ! }{\big(\frac{N-m}{2} \big) ! }.\]
Using  $\langle  M_{m}^{2} \rangle = C_{N,m}^{2}$ we find, 
\[ \langle \Delta M_{m}^{2} \rangle = C_{N,m}^{2} \cos^{2} \Big[  \Big(m+\frac{\chi}{2} Nm \Big) \phi \Big], \]
and
\begin{equation}
	\label{deltax}
	\Delta x \simeq \Delta \phi = \frac{\sqrt{\langle \Delta M_{m}^{2} \rangle}}{|\partial \langle M_{m} \rangle/\partial \phi|} = \frac{2}{ \chi Nm + m}. 
\end{equation}
When $m= N$ and $k=0$, we saturate the quantum Cramer-Rao bound given in Eq.~\eqref{crb1} with $\Delta x \sim 2/\chi N^{2}$ to leading order in $1/N$. If we add photon losses, then the multi-photon coincidence measurements also fail to give super-Heisenberg scaling, and the measurement uncertainty follows the same behaviour as in Fig.~\ref{fig:lossqfi} for the QCRB. \textcolor{black}{It may be noted that alternate strategies like Bayesian estimation protocols~\cite{pezze_mach-zehnder_2008} can be used to saturate the quantum Cramer-Rao bound using measurement operators like $M$ rather than $M_m$. However, this is a multi-round protocol in which the measurement sensitivity improves progressively and each instance using $N$ photons is not always equivalent to the previous one. This make it more difficult to understand clearly the effect of photon loss on the measurement sensitivity while using such protocols. To keep the discussion as clear as possible, we consider only the multi-photon coincidence measurements in the ensuing discussion on finding an optimal initial state that gives the $1/N^2$ scaling even in the presence of photon loss.}

\section{Optimal initial state \label{optimalstate}}

The extreme susceptibility of the enhanced scaling of the measurement uncertainty when using N00N type states leads us to the question of whether an optimal input state that is robust to noise can be found. We start from a general superposition state of the form as input into the interferometer:
\begin{equation}
\left| \psi\right\rangle  = \sum_{k=0}^{\tau} \alpha_k \left\lbrace  \left| N-k, k\right\rangle + \left| k,N-k\right\rangle \right\rbrace   \label{eqn:geninput}
\end{equation}
where $\tau$ is $(N-1)/2 $ if $N$ is odd, $N/2 $ if $N$ is even and the coefficients, $\alpha_i$, are chosen such that the state as a whole is normalised. Note that we have chosen to keep the total photon number fixed for the state at $N$. This places the resource counting at an equal footing between the optimal state and the N00N-type states we previously considered. We are not considering the more general case of a superposition that keeps the mean photon number fixed. 

Under the action of the unitary that corresponds to propagation along the interferometer arms along with the photon loss as described by equation~\eqref{eqn:loss} (considering equal loss in both arms, $\eta_{a_1} = \eta_{a_2} = \eta$), the density matrix of the state of light at the output port takes the following form:
\begin{eqnarray}
	\label{eq:finalst}
	\rho_n & =& \sum_{q,p,k,l}\left[ B^{kl}_{pq}e^{i\frac{\phi}{2} \left[ G(k)-G(l)\right] }\right.  \nonumber \\
	&& \qquad\qquad \times \; \left. \sqrt{P^{kl}_{ppqq}} \left| n^{N-k}_{q},n^{k}_{p}\right\rangle \left\langle n^{N-l}_{q},n^{l}_{p}\right|\right]  \nonumber \\
 	&& +\;  \sum_{q,p,k,l}\left[B^{kl}_{pq}e^{i\frac{\phi}{2}\left[ G(k)+G(l)\right]}\right. \nonumber \\
	&& \qquad\qquad \times \; \left. \sqrt{P^{kl}_{pqqp}} \left| n^{N-k}_{q},n^{k}_{p}\right\rangle \left\langle n^{l}_{q},n^{N-l}_{p}\right| \right] \nonumber \\
	&& + \;  \sum_{q,p,k,l}\left[B^{kl}_{pq}e^{-i\frac{\phi}{2}\left[ G(k)+G(l)\right]}\right. \nonumber \\
	&& \qquad\qquad \times \; \left. \sqrt{P^{kl}_{qppq}} \left| n^{k}_{q},n^{N-k}_{p}\right\rangle \left\langle n^{N-l}_{q},n^{l}_{p}\right| \right] \nonumber  \\
	&& + \; \sum_{q,p,k,l}\left[B^{kl}_{pq}e^{-i\frac{\phi}{2}\left[ G(k)-G(l)\right]}\right. \nonumber \\
	&& \qquad\qquad \times \; \left. \sqrt{P^{kl}_{qqpp}} \left| n^{k}_{q},n^{N-k}_{p}\right\rangle \left\langle n^{l}_{q},n^{N-l}_{p}\right| \right], \qquad
\end{eqnarray}
where $ B^{kl}_{pq} = (\alpha_k\alpha_l^*) B_{pq}$ and 
\[ P^{kl}_{abcd} = P(k,a) P(l,b) P(N-k,c) P(N-l,d).\]
The sums over $q$ and $p$ run from 0 to $k$, $l$, $N-k$ or $N-l$ as the case may be for the states on the right of the sum while $k$ and $l$ run from 0 to $\tau$.

\begin{figure}[!htb]
	\includegraphics[width = 8.5cm]{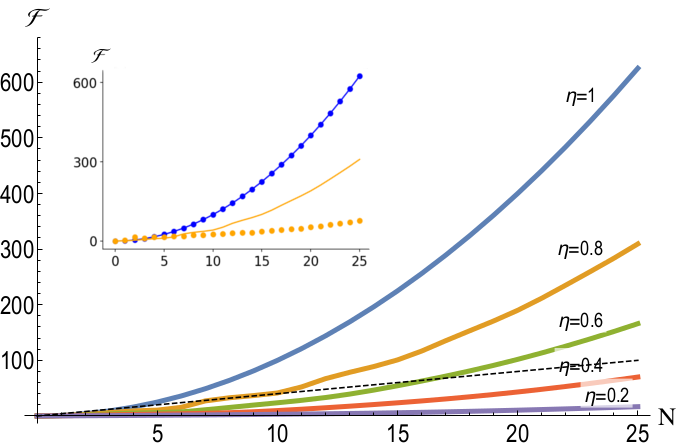}
	\centering
	\caption{Maximum of the quantum Fisher information for the state in Eq.~(\ref{eq:finalst}) vs $N$ varying values of photon losses. Note that $\eta$ is the transmission coefficient of the beam splitters that model the photon loss and so smaller values of $\eta$ correspond to greater photon losses. The Heisenberg limited scaling is the dashed black line for comparison. \textcolor{black}{In the inset, a comparison of performances of the HB state (dotted) and our optimal initial state (solid line) is plotted for $\eta = 1$ (blue) and $\eta = 0.8$ (orange).} }
	\label{fig:qfigen}
\end{figure}

The quantum Fisher information can be computed numerically for the state in Eq.~(\ref{eq:finalst}) corresponding to different values of the photon loss. The coefficients, $\alpha_k$, are then optimised while keeping the state normalised so as to give maximum Fisher information. This optimisation was done for up to $N=25$ and the results are plotted in figure (\ref{fig:qfigen}).
It is clear from Fig.~\ref{fig:qfigen} that the super-Heisenberg scaling for the measurement uncertainty is not fully destroyed in the case of the optimal input state even in the presence of photon loss noise, unlike in the case of the N00N type states. Even with 60\% photon loss, the super-Heisenberg scaling is observed to persist. \textcolor{black}{In the inset plot from Fig.~\ref{fig:qfigen}, we also compare our results with the well known Holland-Burnett (HB) states, which have been shown to attain the Heisenberg limit for phase estimation while maintaining some robustness towards decoherence~\cite{Holland1993,Datta2011}. We see  that while the HB states match the performance of the optimal states that we consider in the zero-loss case, this is not the case when photon losses are present. The scheme} with an optimised superposition of fixed photon number states as input is seen to provide advantageous scaling for the measurement uncertainty as well as much-improved robustness against photon loss.

\begin{figure}[!htb]
	\includegraphics[width = 8.5cm]{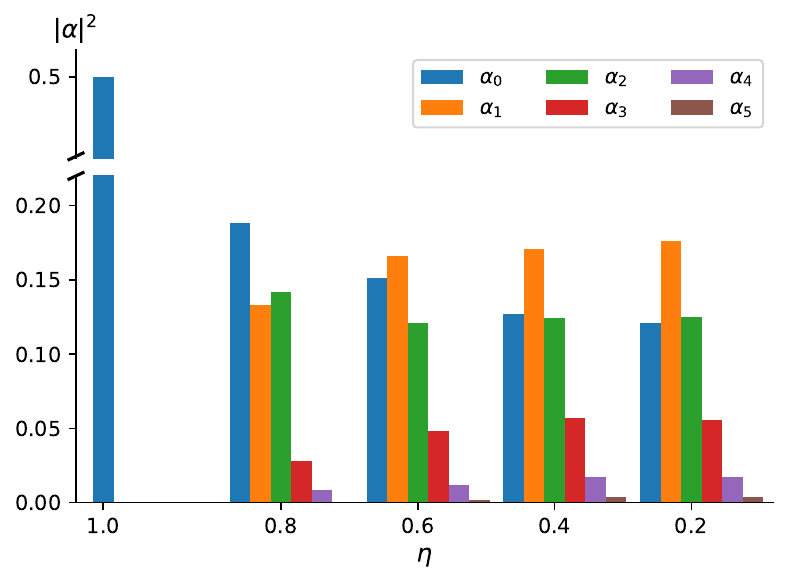}
	\caption{\textcolor{black}{The distributions of the absolute values squares of the first six coefficients, $\alpha_k$, of the optimal state in Eq.~\eqref{eqn:geninput} are shown for different values of $\eta$ for $N=25$. With no losses, only $\alpha_0 \neq 0$ and as the losses increase the the peak of the distribution shifts to higher values of $k$. Note that the y-axis of the figure is broken so as to show all the distributions clearly in the same scale. }} 
	\label{fig:alphas}
\end{figure}

\textcolor{black}{The fixed photon number states in Eq.~\eqref{eqn:geninput} are characterized by the amplitudes $\alpha_k$ that are optimized so as to give the largest possible quantum Fisher information for each value of $\eta$. In Fig.~\ref{fig:alphas}, the distribution of $|\alpha_k|^2$ of the optimal states are shown for different values of $\eta$ with $N=25$. We see that, as expected, when there is no photon loss only $\alpha_0$ is non-zero and the optimal state is the N00N state. When photon losses are present, other values of $\alpha_k$ become non-zero. For $N=25$, there are $13$ different amplitudes defining the state. However, note that even for rather large photon losses, the distribution is limited to small values of $k$ with $|\alpha_k|^2$ for $k>5$ becoming negligible relative to the other amplitudes.}

\subsection{Read-out of the optimal state}

The read-out procedure that would, in practice, saturate the QCRB for the optimal state and lead to super-Heisenberg scalings for the measurement uncertainty even in the presence of noise is theoretically a measurement along the basis given by the eigenstates of the symmetric logarithmic derivative operator in Eq.~\eqref{SLD}. Implementing such measurements is not feasible in practice. The two-photon coincidence measurement of the standard interferometric set up again is not suitable for the optimal $N$ photon state we consider.  We therefore explore the effectiveness of the $m$-photon coincidence measurement from Eq.~\eqref{mphoton} with $m=N$.  

The measurement uncertainty is numerically calculated via error propagation with the measurement operator $M_{N}$ acting on the state in equation~\eqref{eq:finalst} and $\alpha_i$ optimized for maximising the Fisher information given the photon loss. In the absence of photon loss, when the N00N state is used, $\langle M_{N} \rangle$ oscillates sinusoidally. In principle, the measurement uncertainty $\Delta x$ for any value of $\phi$ has to be the same according to Eq.~\eqref{deltax}. However, as can be seen from the same equation, those values of $\phi$ for which $\partial \langle M_{N} \rangle/\partial \phi$ vanishes are not particularly good operating points in real experiments as well as for numerical computation. So we have to minimise the numerically evaluated measurement uncertainty for all values of $\phi \in [0, \pi]$, which makes the computation challenging. 

\begin{figure}[h]
	{\includegraphics[width = 8.5cm]{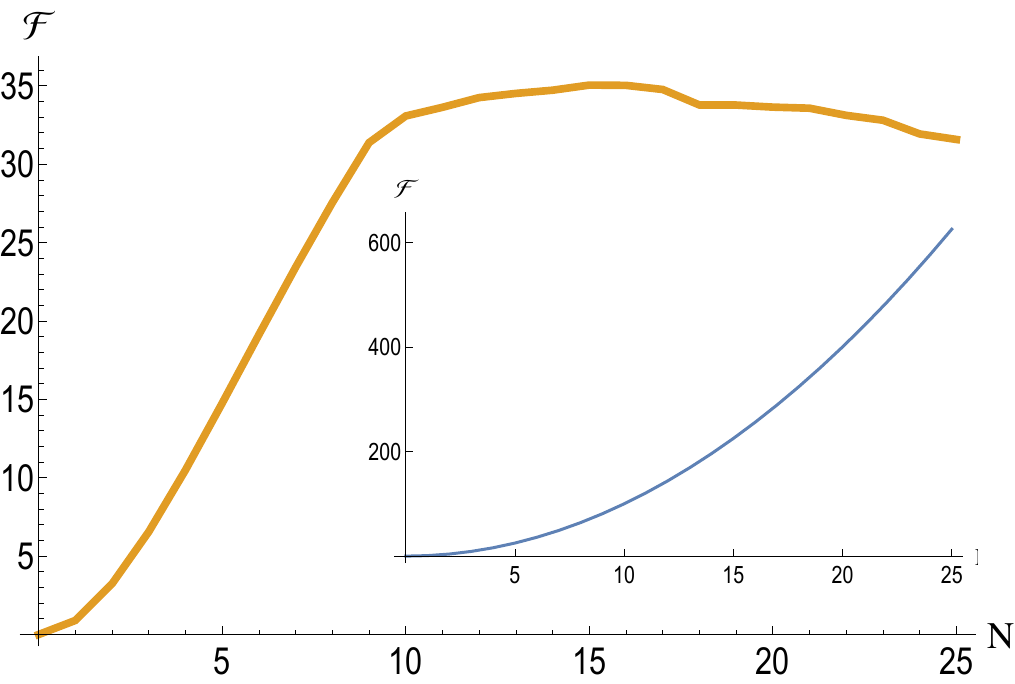}}
	\caption{Inverse of the measurement uncertainty ($1/\Delta x$) vs $N$ for the optimised state in Eq.~(\ref{eq:finalst}) with the  photon loss coefficient set to $\eta = 0.9$. Super-Heisenberg scaling is obtained for small values of $N$ but the curve soon saturates. The inset is the $\eta = 1$ case.}
	\label{fig:readout}
\end{figure}

The inverse of the measurement uncertainty in Fig.~(\ref{fig:readout}) is plotted as a function of $N$ for easy comparison with the QFI. We see that measuring $M_{N}$ can achieve the best scaling offered by the optimised initial state in the absence of any photon loss. Using $M_{N}$ for the read-out is the best choice when N00N states are used as input into the interferometer. The same measurement is not the ideal one for the optimal states that are resilient to photon loss. With only $10\%$ loss, the non linear scaling is very quickly lost and the plot eventually saturates. For very small $N$, the optimal states are relatively close in form to the N00N state and read-out using $M_{N}$ still gives a measurement uncertainty that saturates the QCRB. However, when $N$ increases, this is no longer the case. \textcolor{black}{This behaviour is not completely unexpected because the presence of noise can fundamentally limit the ability to saturate the ultimate super-Heisenberg limit for asymptotically large values of $N$ irrespective of input states or readout methods~\cite{Escher2011}. An alternate measurement strategy can be evolved that maintains nonlinear scaling for higher $N$}. This involves a further optimisation over the measurement operator also which we have not attempted due to the numerical complexity involved. In addition, just like the best possible measurement given by the eigenstates of the symmetric logarithmic derivative operator, the optimised measurement need not be amenable to experimental implementation either. 

\section{Conclusion \label{conclude}}

We have shown that the nonlinear interferometric setup of Luis and Rivas~\cite{Luis2015} can give super-Heisenberg scaling for the measurement uncertainty when using nonclassical states of the N00N type. The quantum Cramer-Rao bound for the interferometric setup was computed when N00N-type, fixed photon-number states were used, and we found that the measurement uncertainty can, in principle, scale as $1/N^{2}$ beating the Heisenberg limited scaling of $1/N$. However, we also found that if the read-out strategy employed on the quantum state of light at the output ports of the interferometer is limited to two-photon coincidences, then the QCRB is not achievable. This brought into focus a potential issue that is often glossed over in the theoretical literature on quantum limited metrology, namely the achievability, in practice of the QCRB. We do find, however, that even with such limitations on the read-out strategy, the nonlinear scheme performs better than the linear one. 

We also addressed the problem of extreme susceptibility of the N00N-type states to photon loss noise. In the presence of even very small amounts of photon loss, we see that the QCRB deteriorates rapidly to scalings that are equal to or even slower than Heisenberg limited scaling. The quantum advantage, as well as the additional advantage provided by the nonlinear coupling, are both found to be extremely fragile with photon-loss. Keeping the overall photon number of the input state fixed, we found an optimal state that provides both super-Heisenberg scaling as well as robustness against photon-loss noise by numerically optimising a generic input state with fixed total photon number. We examined the performance of this state in a practically implementable scheme with $N$-photon coincidences and found that super-Heisenberg scaling can be obtained for low $N$. Finding the optimal read-out strategy that can saturate the QCRB for all $N$ while maintaining the robustness of the scheme against decoherence due to photon loss are avenues to be explored in the future.

Since embedding the interferometer in a nonlinear Kerr media is impractical for LIGO  type interferometers, it would be more  interesting to consider the case where a small section of the interferometer is embedded in such a nonlinear medium. It has been observed that Kerr media and radiation pressure both introduce intensity-dependent nonlinearities~\cite{Loudon1981} in the state of light. By carefully choosing a medium with the appropriate sign and magnitude for its nonlinearity, radiation pressure noise can be compensated for, improving the sensitivity of the interferometer~\cite{Bondu1986}. Combined usage of nonclassical input states along with smaller Kerr cells inside the interferometer arms could, in principle, lead to a viable and practical implementation of a robust, high precision interferometer.

\section*{Acknowledgements}
Anil Shaji acknowledges the support of SERB, Department of Science and Technology, Government of India through grant no. EMR/2016/007221 and the QuEST program of the Department of Science and Technology through project No. Q113 under Theme 4. Both authors acknowledge the centre for high performance computing of IISER TVM for the use of the {\em Padmanabha} cluster. 

\bibliography{NL_biblio}
\nocite{*}

\end{document}